\documentclass{SCGE}
\usepackage[colorlinks,linkcolor=blue,citecolor=blue,urlcolor=blue]{hyperref}

\usepackage[utf8]{inputenc}
\usepackage{rotating}
\usepackage{footnote}
\newcolumntype{C}[1]{>{\centering\let\newline\\\arraybackslash\hspace{0pt}}m{#1}}
\usepackage[math]{blindtext}
\usepackage{times}
\usepackage{newtxtext,newtxmath}
\usepackage{graphicx}	
\usepackage{amsmath}	
\usepackage{amssymb}	
\usepackage{xspace}
\usepackage{bigints}
\usepackage{siunitx}
\usepackage{amsthm}
\usepackage{mathtools}
\usepackage[normalem]{ulem}
\usepackage{float}
\usepackage{color}
\usepackage[sort&compress,numbers]{natbib}
\usepackage{gensymb}

\usepackage{CJK}

\let\citedash\relax
\makeatletter \providecommand{\citedash}{\hbox{-}\penalty\@m}
\makeatother
\begin{document}

\begin{picture}(0,0){\rm
\put(0,-20){\makebox[160truemm][l]{\bf {\sanhao\raisebox{2pt}{.}}
Article  {\sanhao\raisebox{1.5pt}{.}}}}}
\put(0,-34){\jiuwuhao {\textcolor[rgb]{0.5,0.5,0.5}{\sf 
}}}
\end{picture}

\def\bm{\boldsymbol}

\def\dl{\displaystyle}
\def\du{\end{document}}
\def\d{{\rm d}}
\def\e{{\rm e}}
\def\i{{\rm i}}

\Year{2021} %
\Month{XXXX} %
\Vol{xx} %
\No{x} %
\BeginPage{1} %
\AuthorMark{{\rm Ye Feng et al.} }  
\DOI{} 
\ArtNo{000000}

\title[AT2019wey]{Spectral Analysis of New Black Hole Candidate AT2019wey Observed by \emph{NuSTAR}}

\author[1,2]{Ye Feng}{}%
\author[1,2]{Xueshan Zhao}{}
\author[1,2]{Lijun Gou}{}
\author[1,2]{Yufeng Li}{}
\author[3]{James F. Steiner}{}
\author[4,5]{Javier A. {Garc{\'\i}a}}{}
\author[1,2]{Yuan Wang}{}
\author[1,2]{\\Nan Jia}{}
\author[1,2]{Zhenxuan Liao}{}
\author[6]{Huixian Li}{}
\footnote{*Corresponding author: yefeng@nao.cas.cn, xszhao@nao.cas.cn, lgou@nao.cas.cn}
\address[{\rm1}]{Key Laboratory for Computational Astrophysics, National Astronomical Observatories, Chinese Academy of Sciences, Beijing 100012, China;}
\address[{\rm2}]{School of Astronomy and Space Sciences, University of Chinese Academy of Sciences, Beijing 100049, China;}
\address[{\rm3}]{Harvard-Smithsonian Center for Astrophysics, Cambridge, MA 02138, USA;}
\address[{\rm4}]{ahill Centre for Astronomy and Astrophysics, California Institute of Technology, Pasadena, CA 91125, USA;}
\address[{\rm5}]{Dr. Karl Remeis-Observatory and Erlangen Centre for Astroparticle Physics, Bamberg, Germany;}
\address[{\rm6}]{School of Science, Jimei University, Xiamen 361021, China;}

\maketitle \vspace{-3.5mm}{\footnotesize\begin{center} Received Month date, Year; accepted Month date, Year
\end{center}}\vspace*{-5mm}

\begin{center}
\rule{16.5cm}{0.4pt}
\parbox{16.5cm}
{\begin{abstract}
AT2019wey is a new galactic X-ray binary that was first discovered as an optical transient by the Australia Telescope Large Area Survey (ATLAS) on December 7, 2019. AT2019wey consists of a black hole candidate as well as a low-mass companion star ($M_{\text {star }} \lesssim 0.8 M_{\odot}$) and is likely to have a short orbital period ($P_{\text {orb }} \lesssim 8$ h).  Although AT2019wey began activation in the X-ray band during almost the entire outburst on March 8, 2020, it did not enter the soft state during the entire outburst. In this study, we present a detailed spectral analysis of AT2019wey in the low/hard state during its X-ray outburst on the basis of Nuclear Spectroscopic Telescope Array \emph observations. We obtain tight constraints on several of its important physical parameters by applying the State-of-art \texttt{relxill} relativistic reflection model family. In particular, we determine that the measured inner radius of the accretion disk is most likely to have extended to the innermost stable circular orbit (ISCO) radius, i.e., $R_{\text{in}}=1.38^{+0.23}_{-0.16}~R_{\text{ISCO}}$. Hence, assuming $R_{\text{in}}$=$R_{\text{ISCO}}$, we find the spin of AT2019wey to be $a_{*}\sim$ $0.97$, which is close to the extreme and an inner disk inclination angle of ~$i\sim$ $22 ^{\circ}$. Additionally, according to our adopted models, AT2019wey tends to have a relatively high iron abundance of $A_{\mathrm{Fe}}\sim$ 5 $A_{\mathrm{Fe}, \odot}$ and a high disk ionization state of $\log \xi\sim$ 3.4.

\end{abstract}}
\end{center}\vspace*{-0.6cm}

\begin{center}
\parbox{16.5cm}
{\bf\jiuhao \emph{NuSTAR}, black hole physics, X-rays: binaries--stars: individual: AT2019wey}
\end{center}

\begin{center}
{}
\end{center}

\textwidth=178truemm \textheight=236truemm

\wuhao\vspace*{1.5mm}

\renewcommand{\baselinestretch}{1.08} \baselineskip 12.2pt\parindent=10.8pt


\section{Introduction}
\label{section:1}

According to the no-hair theorem of black holes, spin (i.e., angular momentum) is one of the three basic physical quantities of mass, spin, and charge used to determine all properties of a black hole. This parameter is closely related to strong gravity and might be closely associated with jets occurring near the black hole \cite{McClintock2014}. Because the spin of an isolated black hole is difficult to estimate, analysis of the accreting black hole X-ray binary (BHXRB) is thought to be a more appropriate method for studying black hole physics.

In addition to the continuum-fitting method \cite{Zhang1997}, the iron-line profile-fitting method, first proposed by Fabian et al.\cite{Fabian1989} and also known as the $\text {Fe K}  \alpha$ fitting method, is used to measure the spin of a black hole by modeling the reflection component in a spectrum. This method has increased in maturity during the past decades and is widely applied to dozens of both stellar-mass and supermassive black holes \cite{Reynolds2014}. The X-ray reflection spectra include a combination of several components such as the iron fluorescent emission line peak at 6.4--6.97 keV, the absorption K-edge at 7--8 keV, and the Compton hump at 20--40 keV \citep[e.g.,][]{Javier2015}. Among these features, the most prominent is the $\text {Fe K}  \alpha$ emission line. The basis for the $\text {Fe K}  \alpha$ fitting method is to ensure that the inner radius of the accretion disk has extended to the innermost stable circular orbit (ISCO), in short to make sure $R_{\text{in}}$=$R_{\text{ISCO}}$ and that the innermost accretion disk is not fully ionized. By taking advantage of the monotonically increasing relationship between $R_{\text{ISCO}}$ and the black hole spin \cite{Bardeen1972}, $a_{*}$ can be obtained. This approach has been used to measure more than a dozen systems such as XTE J1550-564 \cite{Steiner2011}, GRS 1915+105 \cite{Miller2013}, 4U 1630-472 \cite{King2014}, GS 1354-645 \cite{ElBatal2016}, MAXI J1631-479 \cite{Xu2018}, Swift J1658.2-4242 \cite{Xu2018b}, XTE J1752-223 \cite{Javier2018}, MAXI J1836-194 \cite{Dong2020b}, EXO 1846-031\cite{Draghis2020}, and 4U 1543-47 \cite{Dong2020}.

AT2019wey, also known as SRGA J043520.9+552226 \cite{Cappelluti2011}, ATLAS19bcxp \cite{Tonry2018}, and ZTF19acwrvzk  \cite{Masci2019} and hereafter referred to as AT2019 in the present study, is a Galactic low-mass X-ray binary (LMXB) that was first detected as an optical transient by the Australia Telescope Large Area Survey (ATLAS) \cite{Tonry2018} on December 7, 2019 at the location of $\mathrm{RA}=04^{\text{h}}35^{\text{m}}23.280^{\text{s}}$, $\mathrm{DEC}=+55^{\circ}22^{'}34.25^{''}$ (J2000) \cite{Yao2020}. Since then, comprehensive ultraviolet, optical, near-infrared, radio, and X-ray band observations have been conducted on this source \cite{Yao2020}, \cite{Yao2020b}. The multi-wavelength observations indicated that the compact star of AT2019 is likely a black hole \cite{Yao2020}. In addition, it is possible that this system has a low-mass companion star $M_{\text {star }} \lesssim 0.8 M_{\odot}$ and a short orbital period ($P_{\text {orb }} \lesssim 8$ h) \cite{Yao2020}, \cite{Yao2020b}. Very Long Baseline Array observations also revealed that AT2019 is associated with a steady and compact jet in the low/hard (LH) state \cite{Yadlapalli2021}. However, AT2019 never transitioned into a fully soft state and has remained in the LH state for almost the entire outburst. It is advantageous that the reflection features are much more prominent in the LH state or the intermediate state, which is optimal for the study of the black hole spin using the $\text {Fe K}  \alpha$ fitting method.

In this work, we present a detailed Nuclear Spectroscopic Telescope Array \emph{NuSTAR} spectroscopic analysis for the new black hole candidate AT2019 in the LH state. Using \texttt{relxill}, a new relativistic reflection model, we derive rigorous constraints on the black hole spin as well as the inclination angle of the accretion disk and its states, including the ionization state and iron abundance by applying the X-ray reflection fitting method.

The paper is organized as follows. In Section \ref{section:2}, we describe the data selection and reduction of \emph{NuSTAR} data. In Section \ref{section:3}, we present the detailed spectral analysis and results. A discussion is given in Section \ref{section:4}, and the conclusion is provided in Section \ref{section:5}.

\begin{figure}
\centering
  \includegraphics{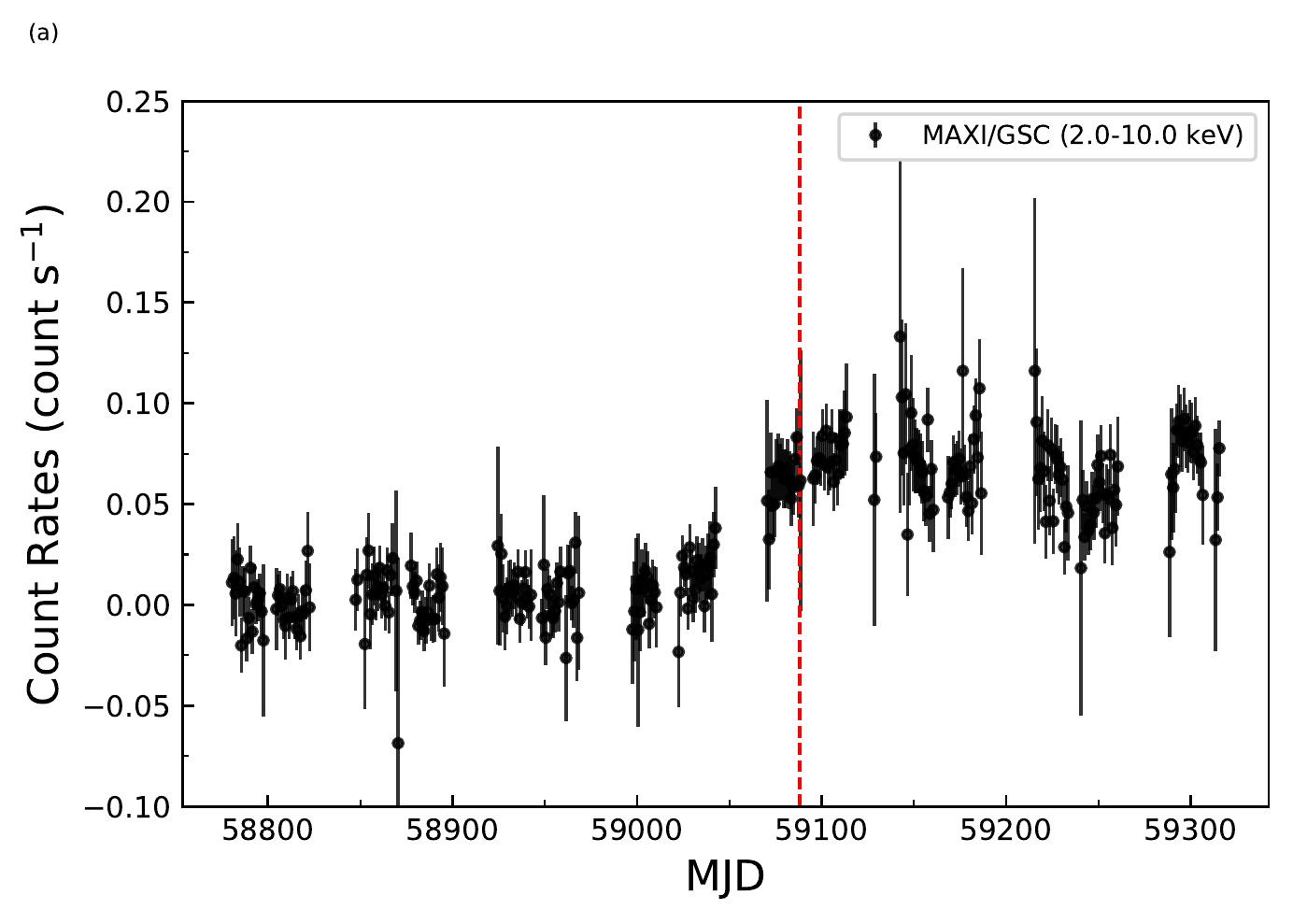}
  \includegraphics{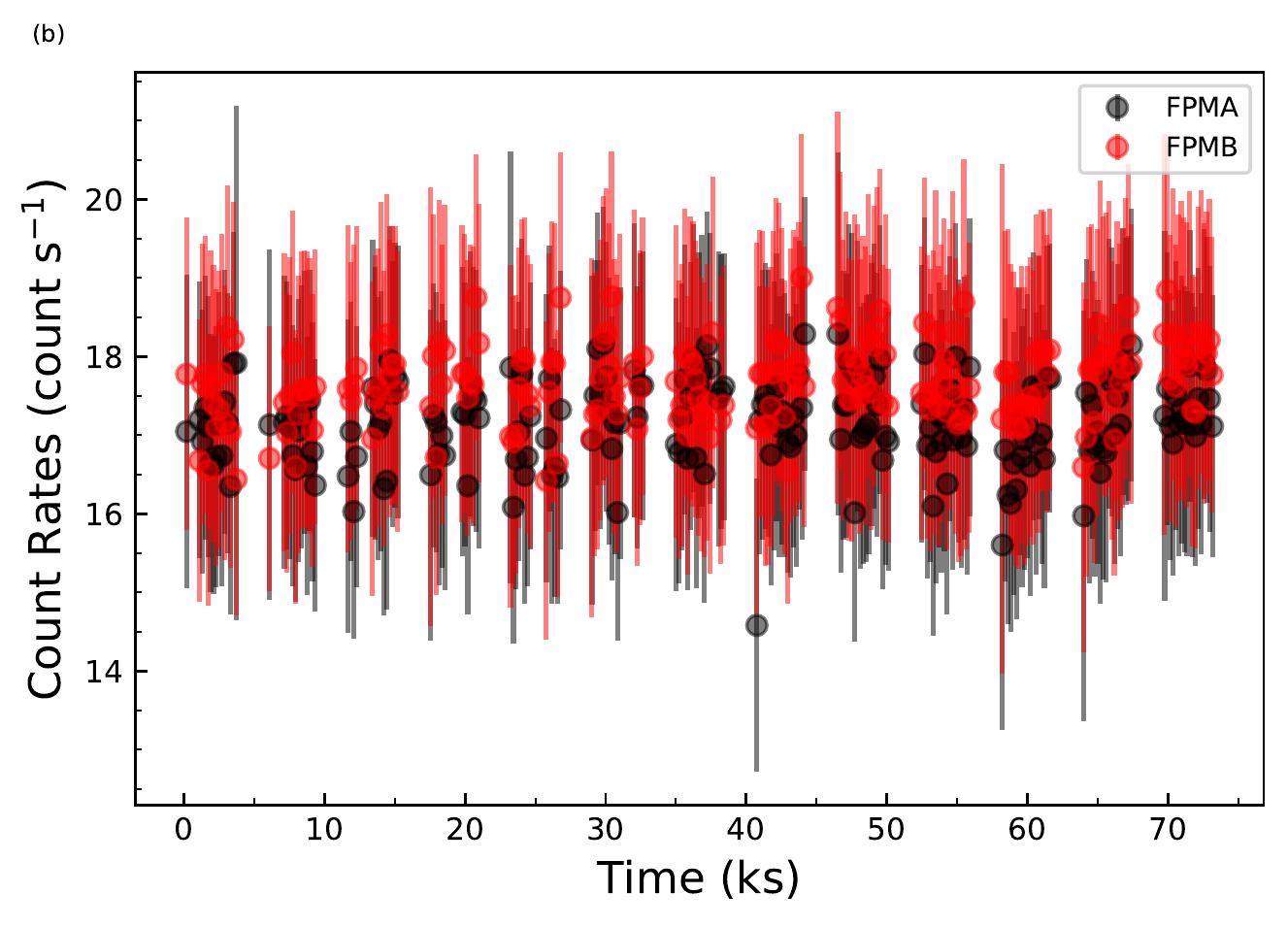}
  \caption{(a) Light curve of AT2019 observed by \emph{MAXI}/GSC at 2.0--10.0 keV. The red dashed line represents the time observed by \emph{NuSTAR} (ObsID: 90601315006). (b) Full band net light curve of AT2019 observed by \emph{NuSTAR} (ObsID: 90601315006). FPMA and FPMB data are plotted in black and red, respectively. The time bins are 20 s in duration.}\label{fig:1}
\end{figure}

\begin{figure}
  \centering
  \includegraphics{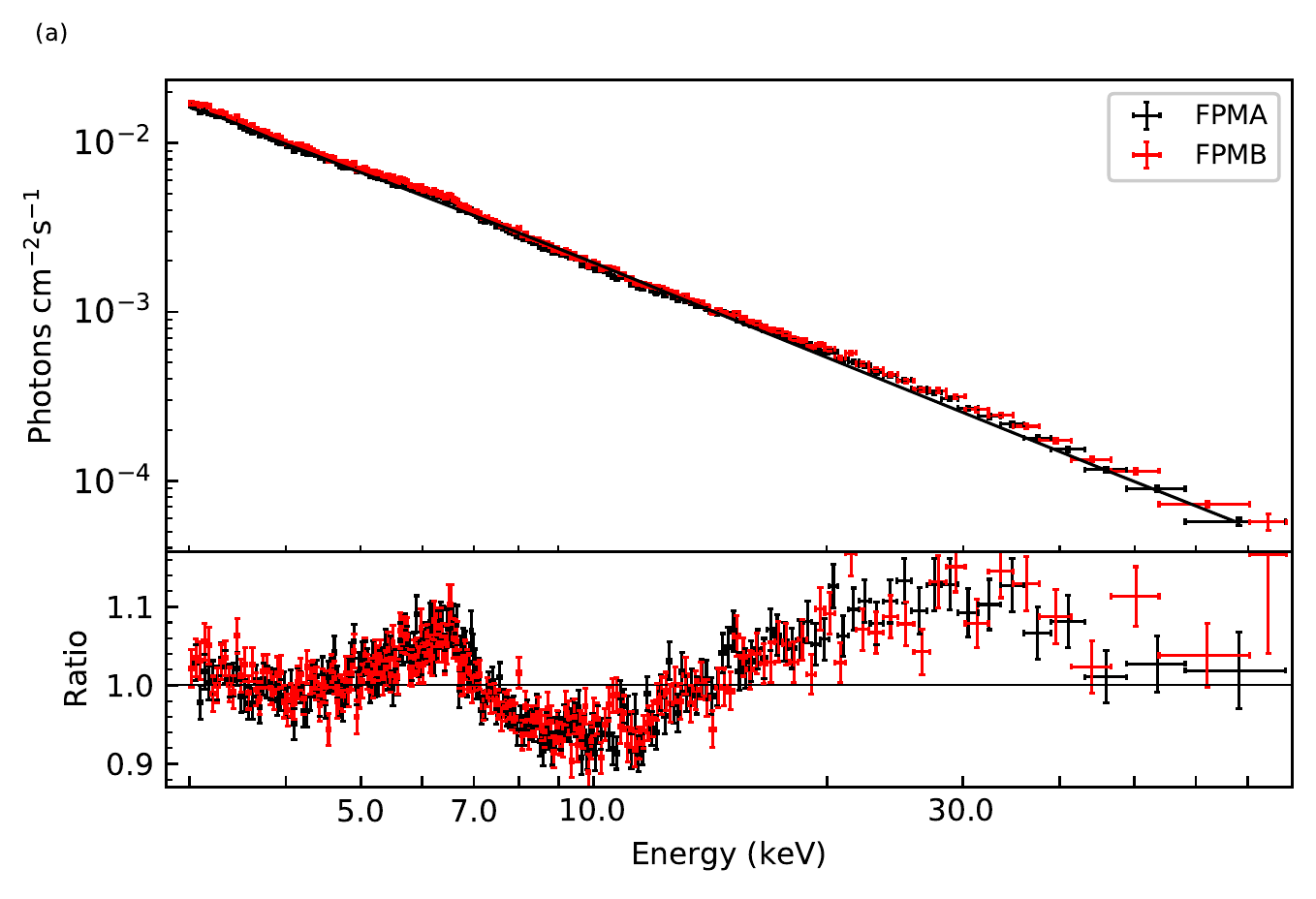}
  \includegraphics{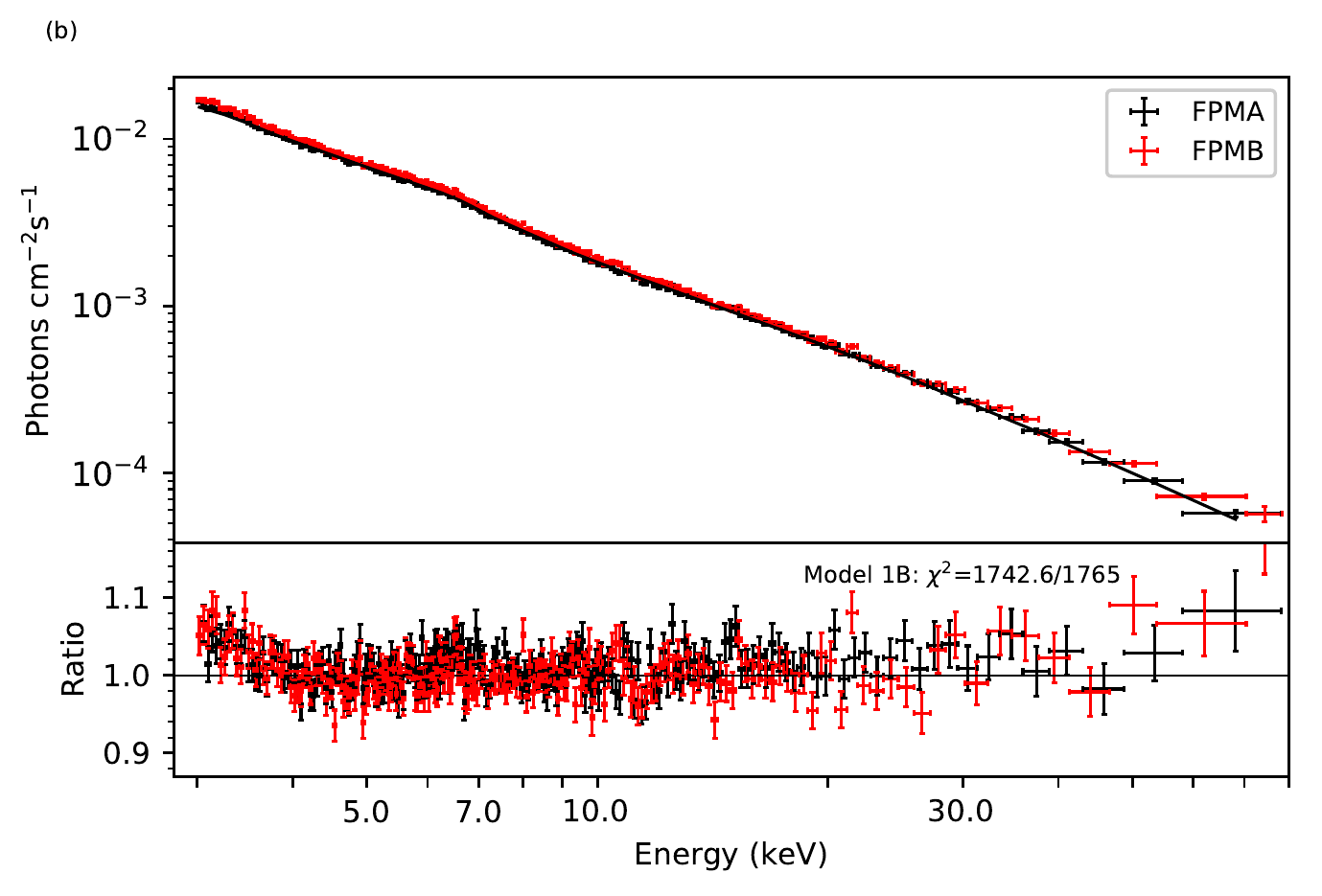}
  \caption{(a) Relativistic reflection features shown as a broadened iron line and Compton hump. The data were fitted over 3.0--79.0 keV, ignoring 6.0--8.0 keV and 20.0--40.0 keV. FPMA and FPMB data are plotted in black and red, respectively. The data have been rebinned for display clarity. (b) Ratio plot of Model 1B. FPMA and FPMB data are plotted in black and red, respectively. The data have been rebinned for visual clarity.}
  \label{fig:2}
\end{figure}

\begin{figure}
  \centering
  \includegraphics[width = 15cm]{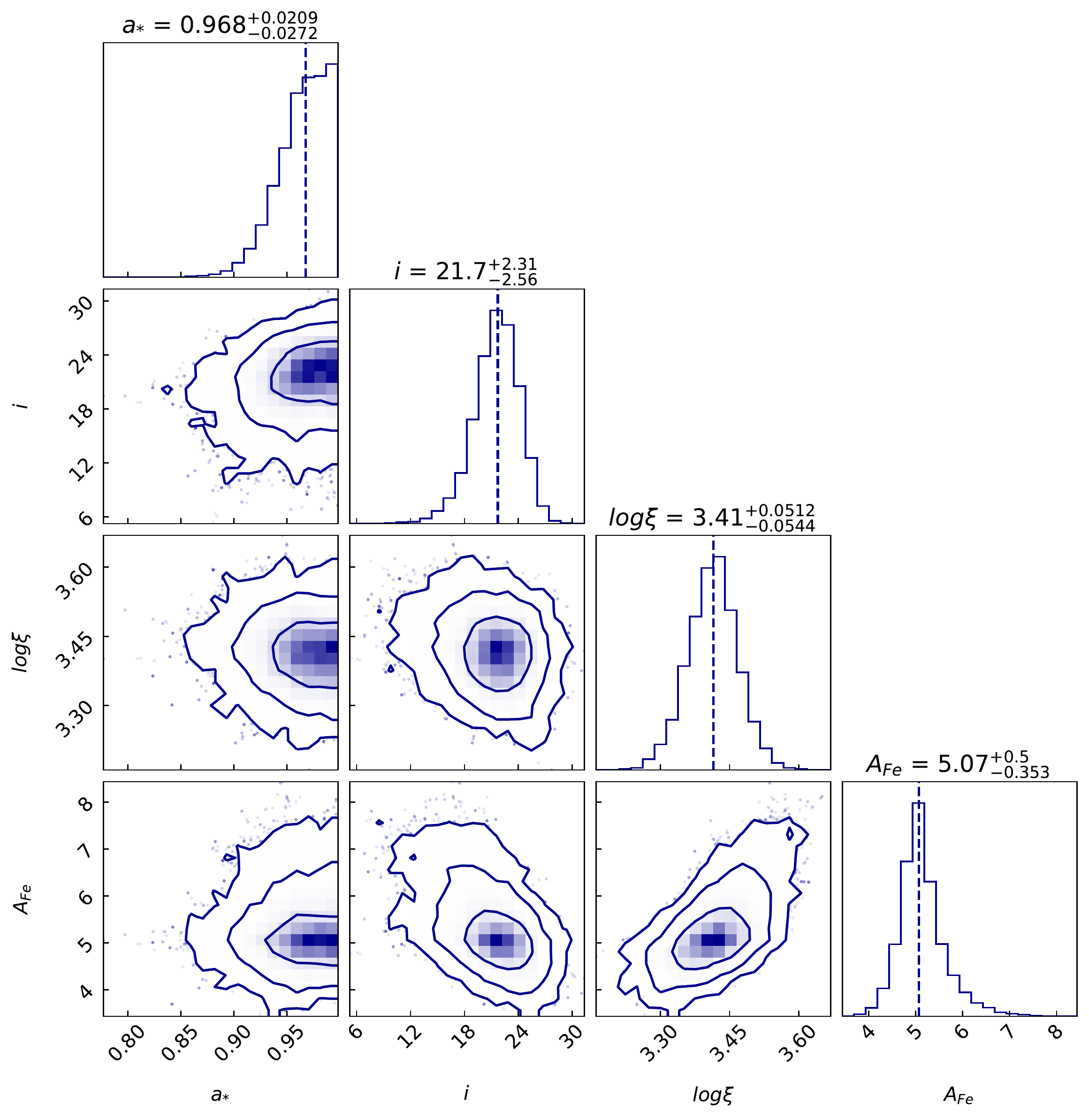}
  \caption{MCMC posterior probability distributions for Model 1B with the \texttt{corner} package for the following parameters: black hole spin (dimensionless unit), inclination angle (in degrees), and ionization state of disk and iron abundance (in units of solar iron abundance). The three contour lines represent $99.7\%$ (3 $\sigma$), $95.4 \%$ ($2 \sigma$), and $68.3 \%$ (1 $\sigma$), respectively. The errors given above the panel are within the $68.3\% $ confidence level.}
  \label{fig:3}
\end{figure}

\begin{figure}
  \centering
  \includegraphics[width = 15cm]{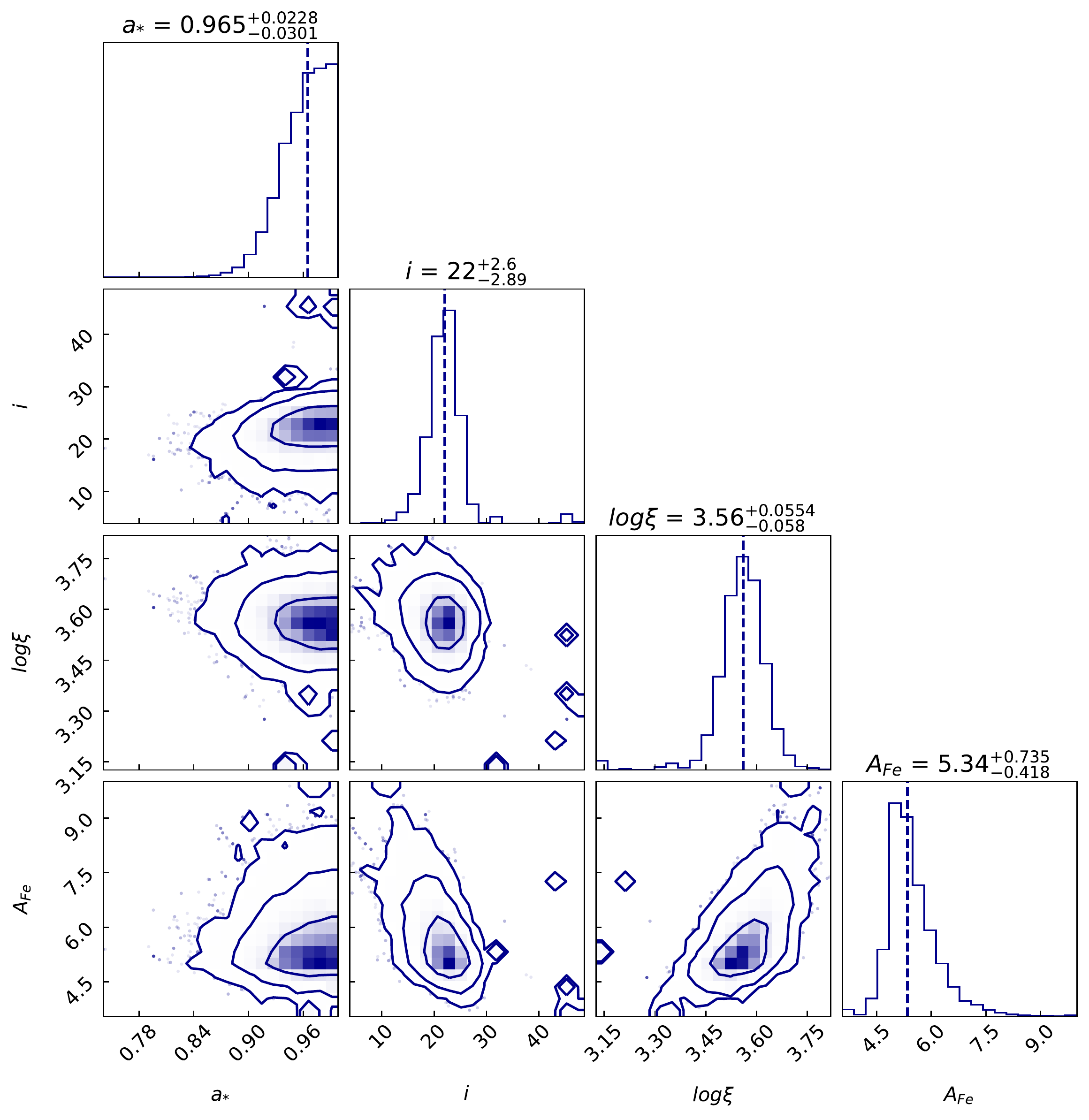}
  \caption{MCMC posterior probability distributions for Model 2, which is a more physical case, with the \texttt{corner} package for the following parameters: black hole spin (dimensionless unit), inclination angle (in degrees), and ionization state of the disk and iron abundance (in units of solar iron abundance). The three contour lines represent $99.7 \%$ (3 $\sigma$), $95.4 \%$ ($2 \sigma$), and $68.3 \%$ (1 $\sigma$), respectively. The errors given above the panel are within the $68.3\% $ confidence level.}
  \label{fig:4}
\end{figure}

\section{DATA SELECTION AND REDUCTION}
\label{section:2}
AT2019 was observed by (\emph{NuSTAR}) \cite{Harrison2013} on August 27, 2020, beginning at 02:51:09 UT for an on-source time of 41.3 ks per detector (ObsID: 90601315006). The deadtime-corrected effective observing time was about 37 ks. The time of this observation is marked by a red dashed line in Figure \ref{fig:1} (a), and its details are listed in Table \ref{table:1}. AT2019 is a faint X-ray source with a flux peak reaching only 20 mCrab in the 0.3--100.0 keV band \cite{Yao2020b}. The net light curve of AT2019 in Figure \ref{fig:1} (b) shows no dips, flares, or other obvious variabilities during this period. We processed the \emph{NuSTAR} data following standard procedures\footnote{\url{https://heasarc.gsfc.nasa.gov/docs/nustar/analysis/nustar_swguide.pdf}} using HEAsoft\footnote{\url{https://heasarc.gsfc.nasa.gov/docs/software/heasoft/download.html}} v2.8 based on the latest calibration files\footnote{\url{https://heasarc.gsfc.nasa.gov/docs/heasarc/caldb/caldb_supported_missions.html}} v20210315. The source spectra were extracted using a circle centered on the source position with a radius of $70^{\prime \prime}$. Using the same radius, the background spectra were extracted from a region as close to the source as possible without including source photons and away from the outer edges of the field of view. Deadtime correction was applied, whereas no multilevel inverter (MLI) correction was required. All data were grouped to achieve at least 30 photons per energy bin using the FTOOLS command \texttt{grppha} \footnote{\url{https://heasarc.gsfc.nasa.gov/ftools/caldb/help/grppha.txt}}. In this work, the spectra of the two \emph{NuSTAR} telescopes, FPMA and FPMB, were fitted jointly over the energy band of 3.0--79.0 keV. Moreover, all spectral analyses in this study were implemented in \texttt{XSPEC v12.11.1}\footnote{\url{https://heasarc.gsfc.nasa.gov/xanadu/xspec/}} \cite{Arnaud1996} with $\chi^2$ statistics. We also ignored the bad channels of all spectra. The calibration of \emph{NuSTAR} has been detailed in Madsen et al.\cite{Madsen2015}. Following the previous absolute flux calibration \cite{Steiner2010}, Crab correction was done in \texttt{XSPEC} using the multiplicative model \texttt{crabcor}. In detail, we fitted the Crab observation of \emph{NuSTAR} for August 29, 2020, in which the date was close to the observation date of August 27, 2020, for the data used in our analysis. Then, we compared the fitting results with the value of Toor and Seward \cite{Toor1974} ($N=9.7 \text{ photons } s^{-1} \text{keV}^{-1}$, $\Gamma=2.1$) which was chosen as a reference. Hence, we obtained a normalization coefficient and a slope difference of $C_{\mathrm{TS}}=0.873$ and $\Delta \Gamma_{\mathrm{TS}}=-0.002$, respectively. Moreover, the limited low-energy band covered by \emph{NuSTAR} leads to poor constraints on the hydrogen column density ($N_{\mathrm{H}}$). Thus, we adopted the estimated result from the $\mathrm{H_{I}}$ Profile Search website\footnote{\url{https://www.astro.uni-bonn.de/hisurvey/profile/index.php}} (i.e., $N_{\mathrm{H}}$ = $3.39 \times10^{21}\mathrm{cm}^{-2}$) and fixed this value in the following work. In addition, we note that this value of $N_{\mathrm{H}}$ is consistent with the result calculated via the HEASARC Tools website\footnote{\url{https://heasarc.gsfc.nasa.gov/cgi-bin/Tools/w3nh/w3nh.pl}}.

\begin{table*}
\caption{\emph{NuSTAR} observation log of AT2019wey\label{table:1}}
\begin{tabular}{ccccccc}
\hline
\hline
ObsID & Module &MJD & Start Time & End Time & Exposure  & Count rates\\
      &        &    &            &          & (ks) & ($\mathrm{cts ~s^{-1}}$)\\
 \hline
 90601315006& FPMA & 	59088.12 	&	2020-08-27 02:51:09	  &	2020-08-27 23:31:09	   &	37.2 	&     11.26 \\

  	  & FPMB & 	 	&	  &		   &	37.1 	&     10.70 \\
  	  \hline
  	  \hline
\end{tabular}
{\\Notes.\\
 In columns 1--7, we display the following information: observation ID (ObsID); module of \emph{NuSTAR}; modified Julian date (MJD); start time; end time; exposure time in units of seconds, which is equal to the dead time-corrected on-source time; and net count rates measured at 3.0--79.0 keV in units of $\mathrm{cts ~s^{-1}}$.
 }
\end{table*}

\begin{table*}\tiny
\caption{Best-fitting parameters of AT2019w in the LH state}\label{table:2}
\begin{tabular}{cccccccccccc}
\hline
\hline
Component  & Parameter & Description & Model 1A & Model 1B &  Model 1C &Model 1D &Model 1E &Model 2 & Model 3& Model 4A& Model 4B\\
& & & free $R_{\text{in}}$ & $R_{\text{in}}$=$R_{\text{ISCO}}$ &$A_{\text{Fe}}$=2.86 &$N_{\text{H}}$=5&free $q_{\text{in}}$&  & & free $\mathrm{log}~(n_{e})$& $\mathrm{log}~(n_{e})=19$ \\
\hline

\tt{constant}& $C_{\mathrm{FPMB}}$ &Cross-normalization &	1.044$\pm$0.002  &1.044 $\pm$0.002&	1.044$\pm$0.002 &1.044$\pm$	0.002  &1.045$\pm$0.002 &  1.044$\pm$0.002&$1.045^{+0.002}_{-0.003}$  &1.045 $\pm$0.002&1.045$\pm$0.002 \\

\tt{TBabs} & $N_{\mathrm{H}}~(\times10^{21}\mathrm{cm}^{-2})$  &Hydrogen column density&$3.39^{*}$ &	$3.39^{*}$ &$3.39^{*}$  &	$5^{*}$ &	$3.39^{*}$  &  $3.39^{*}$ &	$3.39^{*}$  & $3.39^{*}$   &$3.39^{*}$ \\

\tt{relxill(Cp/D)}         &     $a_{*}$                                                                                       &Black hole spin 		            &  $0.998^{*}$                               & $0.97^{+0.02}_{-0.03}$          &$0.95^{+0.02}_{-0.03}$        &$0.97^{+0.02}_{-0.03}$      &$0.97^{+0.02}_{-0.03}$          &  $0.96^{+0.02}_{-0.03}$ 	    &	$0.96^{+0.03}_{-0.04}$        &	$0.97^{+0.02}_{-0.03}$        &	 $0.97^{+0.02}_{-0.03}$\\
\tt{relxill(Cp/D)}         &     $i $~($^{\circ}$)                                                                         &Inclination angle                     &	$22.0^{+2.3}_{-2.4}$                &	$21.7^{+2.3}_{-2.6}$	       &	29.0$\pm$ 1.6              &$20.8^{+2.6}_{-2.9}$           &$24.5^{+4.7}_{-5.3}$             &   $22.0^{+2.6}_{-2.9}$	    &	$17.4^{+4.2}_{-5.4}$            &	$21.7^{+2.3}_{-2.6}$            &	$21.1^{+2.8}_{-3.6}$  \\
\tt{relxill(Cp/D)}         &     $q_{\text{in}}$                                                                             &Emissivity index in the inner region   & $3 ^{*}$                           & $3 ^{*}$                                   & $3 ^{*}$                               & $3 ^{*}$                              &    3.1$\pm$0.2                     &  $3 ^{*}$                                 & $3 ^{*}$                                 &   $3 ^{*}$                                 & $3 ^{*}$\\
\tt{relxill(Cp/D)}         &     $R_{\text{in}} $~($R_{\text{ISCO}}$)                                         &Disk inner radius                    & $1.38^{+0.23}_{-0.16}$              &$1 ^{*}$                                    &$1 ^{*}$                                &$1 ^{*}$                              &   $1 ^{*}$                               &   $1 ^{*}$                               &   $1 ^{*}$                               &    $1 ^{*}$                                &    $1 ^{*}$ \\
\tt{relxill(Cp/D)}         &     $\Gamma$                                                                                 &Photon Index	                    &	1.80$\pm$0.01                         &	1.79$\pm$0.01                 &	1.82$\pm$0.01              &	1.79$\pm$0.01                 &	1.80$\pm$0.01               &  1.82$\pm$0.01                    &	1.79 	$\pm$	0.01            &	1.79$\pm$	0.01            &	1.76 	$\pm$0.01\\
\tt{relxill(Cp/D)}         &     $\log \xi $                                                                                    &Ionization state of disk          &	3.4$\pm$0.1                             &	3.4 	        $\pm$	0.1  &	3.3 	        $\pm$0.1       &	3.5 	        $\pm$	0.1  &  3.4 	$\pm$	0.1         &	3.6	$\pm$	0.1               &	3.6	$\pm$	0.1              &	3.4 	$\pm$	0.1              &	3.5 	$\pm$	 0.1\\
\tt{relxill(Cp/D)}         &     $A_{\mathrm{Fe}}$~($A_{\mathrm{Fe}, \odot}$)                       &Iron abundance                     & $5.1^{+0.5}_{-0.4}$                    & $5.1^{+0.5}_{-0.4}$  	       & $2.86^{*}$                            & $5.2^{+0.5}_{-0.3}$           & $5.1^{+0.6}_{-0.4}$              &  $5.3^{+0.7}_{-0.4}$ 	            &	$5.2^{+0.6}_{-0.3}$              &	$5.1^{+0.5}_{-0.3}$             &	$5.0^{+0.6}_{-0.5}$\\
\tt{relxill(Cp/D)}         &     $\mathrm{log}~(n_{e})$                                                                &Density of disk	                    &	 $15^{\dagger} $                       &	$15^{\dagger} $ 	        &	$15^{\dagger} $             &	$15^{\dagger} $               &  $15^{\dagger} $                    &  $15^{\dagger} $                   &	$15^{\dagger} $                   &	$15.7^{+0.7}_{-0.5}$           &	$19^{*}$\\
\tt{relxill(Cp/D)}         &     $R_{\text{ref}}$                                                                          &Reflection fraction	            &	 $0.38^{+0.03}_{-0.02}$           &	$0.38^{+0.03}_{-0.02}$     &	 0.57$\pm$0.03              &  0.38 	$\pm$0.02               &  $0.40^{+0.05}_{-0.04}$        &  0.34 	$\pm$	0.02     &	0.33$\pm$	0.03            &	$0.38^{+0.03}_{-0.02}$       &	$0.31^{+0.04}_{-0.03}$\\
\tt{relxill(Cp/D)}         &     $N_{\text {relxill~(Cp/D)}}~(\times10^{-3})$                               &Normalization                         &	2.76$\pm$ 0.05                        &	 2.75$\pm$ 0.05 	        &	 2.49$\pm$ 0.04            & 2.72$\pm$ 0.05                & $2.76^{+0.05}_{-0.06}$         &  2.53$\pm$ 0.4  	            &	$2.65^{+0.06}_{-0.07}$        &	2.75$\pm$ 0.05                   &	$2.75^{+0.06}_{-0.09}$\\
\tt{xillver}                   &     $N_{\text {xillver}}~(\times10^{-4}) $                                         &Normalization                         & - 	                                         &  - 	                                          &	-                                     &	-                                       &	-                                      &	-                                             & $1.74^{+0.32}_{-0.35}$         &  -		                           &  -\\
\hline
$\chi_{v}^{2}$                          &  ...                                                                                    &  ...                                          &0.987	                                         &0.987	                                 &1.004               	              &	1.004                                &0.987	                               &	0.988                                    &0.975	                                    &0.988                                      &0.998\\
$\chi^{2} ~(\mathrm{d.o.f.})$   &  ...                                                                                    &  ...                                          &1741.3 (1765) 	                        &1742.6 (1765) 	                        &1774.1 (1766)                        &1772.7 (1765)                   &1742.2 (1764) 	                       &1744.7 (1765)                      &1720.3 (1764) 	                    &1742.6 (1764)                      &1760.8(1765) \\
 \hline
 \hline
\end{tabular}
{\\Notes.\\
Model 1A/B/C/D/E: \texttt{crabcor*constant*TBabs*relxill}; Model 2: \texttt{crabcor*constant*TBabs*relxillCp}; Model 3: \texttt{crabcor*constant*TBabs*(relxill+xillver)}; Model 4A/B: \texttt{crabcor*constant*TBabs*relxillD}. The $\dagger$ marker represents an unchangeable value in these models with a default setting of $n_{e}=10^{15}\text{cm}^{-3}$; the $^{*}$ marker indicates a fixed parameter. The errors were calculated by MCMC at the 68\% confidence interval. Except for Model 1A, all models are under the premise of $R_{\text{in}}$=$R_{\text{ISCO}}$.

}

\end{table*}

\section{SPECTRAL ANALYSIS AND RESULTS}
\label{section:3}
\subsection{Preliminary Spectral Analysis}
\label{section:3.1}
Preliminarily, we fitted the \emph{NuSTAR} data with a simple absorbed power-law model. We ignored the 6.0--8.0 keV and 20.0--40.0 keV regions to demonstrate the reflection features (subgraph (a) of Figure \ref{fig:2}). The residual plot shown in the bottom panel of subgraph (a) in Figure \ref{fig:2}) shows a broadened iron line within the 6.0--7.0 keV energy range, an absorption K-edge at 7.0--8.0 keV, and a Compton hump at 20.0--30.0 keV with $\chi_{v}^{2}=1.65$ (2911.71/1770). During this fit, the photon index $\Gamma$ of the X-ray spectrum, i.e., $N(E) \propto E^{-\Gamma}$, at $\sim$1.8 was within the range of $\sim$ 1.5--2.1, corresponding to a typical LH state as indicated by previous research \cite{mcclintock2003}. Furthermore, owing to the negligible contribution of thermal emission of the accretion disk, as shown in the residual plot of subgraph (a) in Figure \ref{fig:2}, no extra residual except for the iron emission line was obvious at 3.0--10.0 keV. Hence, we did not include the disk component such as \texttt{diskbb} in the subsequent fitting process.

Then, we added a simple Gaussian phenomenological model to fit the iron-line component. The complete model is \texttt{crabcor*constant*TBabs*(powerlaw+gauss)}, in which the \texttt{crabcor} model described in Section \ref{section:2} was used for flux calibration, and the \texttt{constant} model is the energy-independent factor that coordinates the calibration differences between FPMA and FPMB in the joint fit. During the fitting process, we fixed the constant of FPMA $C_{\mathrm{FPMA}}$ to 1 and allowed the constant of FPMB $C_{\mathrm{FPMB}}$ to vary. In addition, \texttt{TBabs} was used as the Galactic inter-stellar medium absorption model, in which we adopted the cross-sections of Verner et al.\cite{Verner1996} and the abundances of Wilms et al.\cite{Wilms2000}. By setting the central energy of the Gaussian line profile to 6.4--6.97 keV, we obtained a broadened Fe line that peaked at $E=6.40\pm 0.07 $ keV with a line width of $\sigma=0.45 \pm 0.04 $ keV and equivalent width of $\sim$110 eV with a reduced chi-square $\chi_{v}^{2}$ value of 2536.02/1767=1.44. This equivalent width is in accordance with the typical value in the LH state \citep[e.g.,][]{Makishima2008}.

\subsection{Relativistic Reflection Model}
\label{section:3.2}
\subsubsection{Incident spectrum with a high-energy cutoff power-law}
\label{section:3.2.1}
To measure the spin of AT2019, we modeled the reflection component using \texttt{relxill}\footnote{\url{http://www.sternwarte.uni-erlangen.de/~dauser/research/relxill/}} v1.4.3 \citep[]{Javier2014, Dauser2014}, which is a sophisticated relativistic reflection model that combines two models. The first, \texttt{relline}, considers the relativistic broadening effect \citep[]{Dauser2010, Dauser2013}, and the second, \texttt{xillver}, can produce the reflection spectrum for an ionized accretion disk without considering the relativistic effect, given a cutoff power-law incident spectrum~\citep[]{Javier2010, Javier2011, Javier2013}. This model can be expressed as \texttt{crabcor*constant*TBabs*relxill} (Model 1).

It is worth noting that in the LH state, BHXRB is usually thought to have an advection-dominated accretion flow with a truncated inner disk (Figure 1 in Esin et al.\cite{Esin1997}). As previously discussed in Section \ref{section:1}, ensuring that the inner radius of the accretion disk extends to ISCO is key for estimating the black hole spin. For this reason, we first checked whether the inner radius was truncated using the approach suggested in Garc{\'\i}a et al.\cite{Javier2018}. We fixed the dimensionless spin of the black hole ($a_{*}$) to its maximum value of 0.998 in Model \texttt{relxill} and allowed the inner radius ($R_{\text{in}}$) to be fitted freely. Because the emissivity index q of the reflection emissivity profile $\epsilon~(r) \propto r^{-q}$ is difficult to be constrained in most cases, we followed the previous work and assumed $q_{\text{in}}=q_{\text{out}}=3$ \citep[e.g.,][]{Novikov1973}, where $q_{\text{in}}$  and  $q_{\text{out}}$ are the emissivity index in the inner and outer regions, respectively. The spectra did not show obvious cutoff; therefore, the cutoff energy ($E_{\text{cut}}$) cannot be accurately constrained owing to the limited energy coverage of \emph{NuSTAR}. Accordingly, we fixed the cutoff energy at a typical and reasonable value of $E_{\text{cut}}$ = 300 keV following the traditional case \citep[]{Javier2015, Javier2010, Ross2005}. The outer radius of disk ($R_{\text{out}}$) was fixed at its default value of 400 $R_{\text{g}}$, where $R_{\mathrm{g}} = \mathrm{GM} / c^{2}$ is the gravitational radius. Furthermore, as a source within our Galaxy, the redshift (z) of AT2019 was fixed at zero. Other parameters were set to vary freely including the inclination angle ($i$), photon index ($\Gamma$), ionization state of the accretion disk ($\log \xi $), iron abundance ($A_{\mathrm{Fe}}$), reflection fraction ($R_{\text{ref}}$), and normalization of \texttt{relxill} ($N_{\text {relxill}}$).

We defined the free $R_{\text{in}}$ model as Model 1A, the fitting results of which are listed in Table \ref{table:2}. The inner radius of the accretion disk was $1.38^{+0.23}_{-0.16}~R_{\text{ISCO}}$, which is close to the ISCO with a reduced chi-square $\chi_{v}^{2}$ value of 0.987 (1741.3/1765) and suggests that the inner disk is not truncated. Hence, the hypothesis of $R_{\text{in}}$=$R_{\text{ISCO}}$ is reasonable, and we assumed that the inner radius of disk $R_{\text{in}}$ was equal to $R_{\text{ISCO}}$ in the subsequent fits. In addition to allowing the spin parameter to be fitted freely, the other parameters were the same as those set in the previous model. To easily distinguish the models, that with the fixed $R_{\text{in}}$ was defined as Model 1B, which we regarded as our benchmark model. From Table \ref{table:2} and subgraph (b) of Figure \ref{fig:2}, we can readily determine that Model 1B is also able to fit the spectra well. In this case, we obtained a black hole spin $a_{*} of 0.97^{+0.02}_{-0.03}$ and an inclination angle $i of 21.7^{+2.3}_{-2.6}$ with a $\chi_{v}^{2}$ value of 0.987 (1742.6/1765) within the $68.3\% $ confidence level ($1\sigma$).

\subsubsection{Incident spectrum with the Comptonization continuum}
\label{section:3.2.2}
Next, to consider a more physical Comptonization continuum case, we replaced \texttt{relxill} with \texttt{relxillCp}, which can be expressed as \texttt{crabcor*constant*TBabs*relxillCp} (Model 2). The distinction between these two models (\texttt{relxill} versus \texttt{relxillCp}) is that the incident spectrum of the former is assumed to be a cutoff power-law, whereas the latter is an Comptonized continuum (Cp) represented by the \texttt{nthcomp} model. Hence, the parameter $E_{\mathrm{cut}}$ in \texttt{relxill} was replaced by the electron temperature ($k T_{\text{e}}$) of the corona in \texttt{relxillCp}. According to $k T_{\text{e}}= 1/3~E_{\mathrm{cut}}$ ($1/2~E_{\mathrm{cut}}$) \cite{Petrucci2001}, we fixed $k T_{\text{e}}$ at one-third the default value of $E_{\mathrm{cut}}$ (i.e., $k T_{\text{e}}=100 $ keV) as in most relevant cases \citep[e.g.,][]{Makishima2008, Poutanen2009, Malzac2009, Kolehmainen2014}. Other parameters remained the same as those in Model 1B.

\subsubsection{Additional Non-Relativistic Reflection Model \texttt{xillver}}
\label{section:3.2.3}
Although we found no obvious narrow iron-line features in the residual plots of models 1 and 2, many previous studies have shown that a cloud far from the central black hole can also produce a narrow line feature \citep[e.g.,][]{Javier2015b, Tomsick2018, You2021} that can be modeled with a non-relativistic reflection component. Therefore, we further investigated the effect by including a second non-relativistic reflection component. Accordingly, we added another reflection model, \texttt{xillver}, defined as Model 3, that can be expressed as \texttt{crabcor*constant*TBabs*(relxill+xillver)}. Except for the ionization state of the accretion disk ($\log \xi $), the reflection fraction ($R_{\text{ref}}$), and the normalization of \texttt{xillver} ($N_{\text {xillver}}$), we linked the parameters of model \texttt{xillver} with that of model \texttt{relxill}. Furthermore, we fixed the $\log \xi =0$ in \texttt{xillver}, which means that the materials are neutral owing to the farther distance from the central black hole. Moreover, we fixed the reflection fraction of \texttt{xillver} to -1 and let $N_{\text {xillver}}$ be free.

\subsubsection{Markov Chain Monte Carlo Analysis}
\label{section:3.2.4}
After fitting the spectra with each model listed above, we conducted a more robust analysis for each model via Markov Chain Monte Carlo (MCMC) in \texttt{XSPEC}. The \texttt{EMCEE} implemented in \texttt{XSPEC} was provided by A. Zoghbi (\url{https://github.com/zoghbi-a/xspec_emcee}). For each individual spectrum, we adopted the Goodman--Weare algorithm \cite{goodman2010} with 100 walkers and a total chain length of $10^{5}$ steps. We discarded the first $10^{4}$ steps in the burn-in phase for each chain. The convergence was evaluated by requiring the average walker to traverse at least ten autocorrelation lengths for each MCMC parameter. In addition, we created time-series plots for all free parameters in each model to check the convergence. The best-fitting results of each model obtained by MCMC are listed in Table \ref{table:2}. It should be noted that all errors were calculated with a 68.3\% confidence level. Moreover, because MCMC does not optimize the reduced chi-square, the best fit values ($\chi_{v}^{2}$) were obtained from \texttt{XSPEC} whereas errors were estimated with MCMC. As shown in Figure \ref{fig:3}, we used the \texttt{corner}\footnote{\url{https://corner.readthedocs.io/en/latest/install.html}} package to create the MCMC contour maps of several parameters, including black hole spin, inclination angle, ionization state of the disk, and iron abundance, which were used for Model 1B. Obviously, all of these parameters can be well constrained with slight degeneracy. In Figure \ref{fig:4}, we show the MCMC contour maps of Model 2, which corresponds to a more physical situation.

\section{DISCUSSION}
\label{section:4}
In this work, we analyzed the \emph{NuSTAR} broadband X-ray data of AT2019 in the LH state. Our preliminary analysis showed that the estimated inner radius was likely close to the ISCO radius, i.e., $R_{\text{in}}=1.38^{+0.23}_{-0.16}~R_{\text{ISCO}}$ during the observation, which ensured the reliable application of the X-ray reflection fitting method. We applied the three different models mentioned in Section \ref{section:3} (model 1A/B, 2, and 3) to the same spectra over 3.0--79.0 keV. A comparison of models 3 and 1B revealed that the addition of the non-relativistic reflection components did not improve the fitting result and led to a smaller inclination value of $17\degree.4_{-5.4}^{+4.2}$. Thus, we determined that the \texttt{xillver} component is not necessary for AT2019. For all three models, according to the best-fitting results shown in Table \ref{table:2}, the reduced chi-square values were close to 1, and no obvious residual feature was detected in the ratio plot, as shown in subgraph (b) of Figure \ref{fig:2}. Moreover, the results among these three different models were consistent. The black hole spin ranged between $\sim$ $0.96$ and $0.97$, which is close to an extreme value. In addition, the inclination angle of the inner disk was $< 30 ^{\circ}$, and the ionization and abundance of the accretion disk were $\sim3.4$--$3.6 $ and $\sim 5.1$--$5.3 A_{\mathrm{Fe}, \odot}$, respectively.

\subsection{Effect of Iron Abundance}
Obviously, Model 1A, Model 1B, Model 2, and Model 3 in our analysis tended to have super-solar iron abundances of $5.1^{+0.5}_{-0.4}$, $5.1^{+0.5}_{-0.4}$, $5.3^{+0.7}_{-0.4}$, and $5.2^{+0.6}_{-0.3}$, respectively. However, Yao et al.\cite{Yao2020b} adopted different models and obtained a relatively lower value of $\sim$ 2.86 for the iron abundance. In addition, several sources have shown degeneracy between the black hole spin and iron abundance \citep[e.g.,][]{Reynolds2012, Steiner2012}. Considering the influence of iron abundance on other parameters, we fixed the iron abundance at $A_{\mathrm{Fe}}$ = 2.86 $A_{\mathrm{Fe}, \odot}$ (Model 1C), following Yao et al.\cite{Yao2020b}. For the convenience of comparison, we also listed the fitting results of Model 1C in Table \ref{table:2}. Compared with Model 1B, the $\Gamma$ increased from 1.79 to 1.82, the spin decreased from 0.97 to 0.95, the ionization state of the disk dropped from 3.4 to 3.3. However, the inclination angle changed from 21.7$^{\circ}$ to 29.0$^{\circ}$, showing an increase of 34$\%$; this result is consistent with the fitting results obtained by Yao et al.\cite{Yao2020b} for different models. In addition, the reduced chi-square $\chi_{v}^{2}$ increased from 0.987 to 1.004. Thus, with the exception of the inclination angle, fixing $A_{\mathrm{Fe}}$ to 2.86 $A_{\mathrm{Fe}, \odot}$ had little impact on the parameters, particularly the spin.

Prior to this work, several sources fitted with model \texttt{relxill} have shown large iron abundance such as 4U 1543-47 \cite{Dong2020b}, MAXI J1836-194 \cite{Dong2020}, GX 339-4 e.g., \cite{Javier2015b}, Cyg X-1 e.g., \cite{Parker2015}, and V404 Cygni \cite{Walton2017}. This situation also occurs in active Galactic nucleus (AGN) systems \cite{Dauser2012}. For the \texttt{relxill(Cp)} model, the default value of the electron density of the accretion disk ($n_{e}$) is usually set at a relatively low level of $\sim 10^{15}\text{cm}^{-3}$, whereas the predicted electron density of an actual black hole accretion disk in a series of theories is significantly higher \citep[e.g.,][]{Noble2010, Schnittman2013}. When $n_{e}$ increases, the main effect on the spectrum is a significant increase in the flux at a low-energy band (Figure 4 in Garc{\'\i}a et al.\cite{Javier2016} and Figure 7 in Tomsick et al.\cite{Tomsick2018}). Although the effects mentioned above have no direct influence on the profile of the iron line, the dramatic change at the low-energy band of the spectrum has an indirect influence on the iron abundance value measured by spectral fitting, which can artificially introduce a large iron abundance. In the actual fit to MAXI J1836-194, Dong et al.\cite{Dong2020} confirmed that the value of iron abundance decreased when the electron density increased.

A high-density version of \texttt{relxill} known as \texttt{relxillD} is available that can achieve a maximum electron density $n_{e}$ of $10^{19}\text{cm}^{-3}$. Therefore, \texttt{relxillD} was also applied to fit the spectra to further identify the origin of the super-solar iron abundance, which is model selection in the fit or intrinsic property of the accretion disk. In this case, the complete model can be expressed as \texttt{crabcor*constant*TBabs*relxillD} (Model 4). We attempted two fitting cases in which the electron density $\mathrm{log}~(n_{e})$ varied freely (Model 4A) and the $\mathrm{log}~(n_{e})$ was fixed at the upper limit of $19$ (Model 4B). The other parameters remained the same. For convenient comparison, the fitting results of Model 4 are also listed in Table \ref{table:2}. In this model, the electron density $n_{e}$ obtained by MCMC remained about $\sim 10^{15}\text{cm}^{-3}$ which is consistent with that in model 1--3. In addition, other parameters such as spin, inclination angle, ionization state, and iron abundance of the disk were the same as in Model 1B. The photon index showed a slight difference between that in Model 4B ($\Gamma=1.76\pm 0.01$) and Model 1B ($\Gamma =1.79\pm 0.01$). Although there is no obvious change was noted in other parameters, we were more inclined to apply the fitting results of Model 4A, where $\mathrm{log}~(n_{e})$ is free. We concluded that $\mathrm{log}~(n_{e})$ has no significant effect on the parameter $A_{\mathrm{Fe}}$ for AT2019. Moreover, according to the histogram distribution of $A_{\mathrm{Fe}}$ obtained by modeling the reflection spectra for both 13 AGN and 9 BHBs of Garc{\'\i}a et al.\cite{Javier2018b}, as shown in Figure 3 of that study, the iron abundance peak at about $ 5$ appears to be a typical value for most sources. In addition, several studies reported that the super-solar iron abundances might be intrinsic \citep[e.g.,][]{Tomsick2018, Reynolds2012, Wang2012}. Therefore, according to the models used in the present study, we concluded that AT2019 is likely to have a relatively large iron abundance of $\sim5$.

\subsection{Effect of Hydrogen Column Density}
\label{section:4.2}
According to the extinction law, Yao et al.\cite{Yao2020b} deduced a hydrogen column density range of $4.4<N_{\mathrm{H}}<6.7$ in units of $10^{21}\mathrm{cm}^{-2}$. Here, we investigated whether the hydrogen column density influenced our fitting results. We took the median value $N_{\mathrm{H}}=5\times10^{21}\mathrm{cm}^{-2}$ of the result obtained by Yao et al.\cite{Yao2020b} as an example to show the effect of changing the $N_{\mathrm{H}}$, as shown in the Model 1D entry in Table \ref{table:2}. In contrast, when $N_{\mathrm{H}}$ changed from $N_{\mathrm{H}}=3.39\times10^{21}\mathrm{cm}^{-2}$ in Model 1B to $N_{\mathrm{H}}=5\times10^{21}\mathrm{cm}^{-2}$ in Model 1D, the median value of the black hole spin remained unchanged, whereas the reduced chi-square $\chi_{v}^{2}$ increased from 0.987 to 1.004. As expected, a slight increase in $N_{\mathrm{H}}$ had a negligible effect on the spin of AT2019.

\subsection{Effect of Emissivity Index}
\label{section:4.3}
The emissivity index q will affect the spin measurement \citep[e.g.,][]{Duro2016, Javier2018}. For this reason, we conducted further analysis, assuming that the emissivity index can be described by a cutoff power-law. Specifically, the emissivity index in the inner region ($q_{\text{in}}$) was set to be free, as defined in Model 1E. Except for $q_{\text{in}}$, the settings of the parameters were consistent with those in Model 1B. To facilitate comparison, the fitting results are also listed in Table \ref{table:2}. At $\chi_{\nu}^{2}$=1742.2/1764=0.987, the reduced chi-square remained almost the same at that in Model 1B, at $\chi_{\nu}^{2}$=1742.6/1765=0.987. In addition, the best-fitting parameters did not change significantly, as indicated by the spin of $a_{*}=0.97^{+0.02}_{-0.03}$ and inclination angle of $i= 24\degree.5^{+4.7}_{-5.3}$. Therefore, considering the principle of reducing the complexity of the model as much as possible during the fitting process and combining the fitting results of Model 1E, we conclude that the fitting results of Model 1B are reasonable.

\subsection{Possible Reason for Low Reflection Fraction}
\label{section:4.4}
The reflection fraction ($R_{\text{ref}}$) is defined as the ratio of intensity emitted toward the disk compared with those escaping to infinity \cite{Dauser2016}. The four different models in nine different forms all showed low reflection fractions of $R_{\text{ref}}$=$0.39^{+0.03}_{-0.02}$ for Model 1A, $0.38^{+0.03}_{-0.02}$  for Model 1B, 0.57 $\pm$ 0.03 for Model 1C, 0.38 $\pm$ 0.02 for Model 1D, $0.40^{+0.05}_{-0.04}$ for Model 1E, 0.34 $\pm$ 0.02 for Model 2, 0.33 $\pm$ 0.03 for Model 3, $0.38^{+0.03}_{-0.02}$ for Model 4A, and $R_{\text{ref}}$=$0.31^{+0.04}_{-0.03}$ for Model 4B. Although the $R_{\text{ref}}$ for Model 1C increased by 0.19 relative to that in Model 1B, the value is still less than 1. According to the results of Beloborodov \cite{Beloborodov1999}, a mildly relativistic outflowing corona leads to aberration, which decreases the X-ray emission toward the disk and reduces the reprocessed radiation from the disk and spectral hardening of the BHXRBs. This interaction can be described as $\Gamma \approx 1.9 / \sqrt{B}$, where $B=\gamma(1+\beta)$ is the aberration factor for the corona with a bulk velocity $\beta=v/c$, and $\gamma$ is the Lorentz factor $\gamma=1 / \sqrt{1-\beta^{2}}$. Therefore, for AT2019, the observed photon index $\Gamma \sim 1.8$ in the LH state and the low reflection fraction are both related to an outflowing corona with a bulk velocity $\beta \sim 0.6$.

As reported by Steiner et al.\citep[]{Steiner2016, Steiner2017}, other explanations are possible for $R_{\text{ref}}$$~\textless$ 1. In the hard state, a black hole has a thick corona that intercepts most thermal photons and Compton-scatters them. Similarly, reflection emission from the disk will be Compton-scattered by the corona. This process can dilute the emergent reflection signature to exp(-$\tau$) and convert the scattered reflection emission into a power-law continuum. That is, most hard state reflection emissions should be Comptonized. In Feng et al. (in preparation), we provide a specific discussion on this explanation.

\subsection{Possible Origin of High Spin}
\label{section:4.5}
Most observed black holes have positive spins. Nielsen\cite{Nielsen2016} showed that the distribution of these black holes with prograde spins is not completely random and appears to follow a particular pattern. High-mass X-ray binaries (HMXBs) usually have higher spins, whereas those of LMXBs range from high spins to moderate and slow spins. The spin distributions differ completely among the BHXRB types. Except for the case in which the sample of the black hole is too small, this distribution could be more closely related to the origin of the spin. It is widely accepted that HXMBs have difficulty in acquiring relatively higher spins through wind accretion owing to the limited lifetime of the companion star and the relatively lower accretion efficiency of the stellar wind. Axelsson et al.\cite{Axelsson2011} reported that the higher spins of HMXBs are caused mainly by the collapse of the progenitor stars of black holes, whereas part of the high spins in LXMBs is caused by the long-term stable accretion of black holes \cite{Fragos2015}. However, M{\'e}ndez et al.\cite{Moreno2008, Moreno2011} determined that the high spin of two HMXBs, M33 X-7 and LMC X-1, originated from super-Eddington accretion. However, companion star of AT2019,  has a very small mass $M_{\text {star }} \lesssim 0.8 M_{\odot}$. Thus, even if the black hole accretes the entire mass of the companion star, the amount of spin change would be minimal, as shown in Figure 3 of King and Kolb \cite{King1999}. Therefore, we inferred a natal origin for its spin.

\section{CONCLUSION}
\label{section:5}
In this study, we conducted a detailed spectroscopic analysis on a clear relativistic reflection feature detected in the new black hole candidate AT2019 observed by \emph{NuSTAR} in the LH state. During this observation, the inner radius of the disk extended to the ISCO, which ensured that the disk was not truncated and enabled the use of the X-ray reflection fitting method to measure the spin. We jointly fitted the FPMA and FPMB data with the sophisticated relativistic reflection model \texttt{relxill} family, and we discussed an additional non-relativistic case utilizing the \texttt{xillver} model. According to Model 2, which is a more physical case, we obtained the following results: spin of AT2019 $a_{*}: 0.96^{+0.02}_{-0.03}$, inclination angle $i: 22\degree.0^{+2.6}_{-2.9}$, iron abundance $A_{\mathrm{Fe}}: 5.3^{+0.7}_{-0.4}$, and ionization state of disk $\log \xi$: 3.6 $\pm$	 0.1. In addition, all models tested in this study reached consistent conclusions of $a_{*}=0.97^{+0.02}_{-0.03}$ for Model 1B, $a_{*}=0.95^{+0.02}_{-0.03}$ for Model 1C, $a_{*}=0.97^{+0.02}_{-0.03}$ for Model 1D, $a_{*}=0.97^{+0.02}_{-0.03}$ for Model 1E, $a_{*}=0.96^{+0.02}_{-0.03}$ for Model 2, $a_{*}=0.96^{+0.03}_{-0.04}$ for Model 3, $a_{*}=0.97^{+0.02}_{-0.03}$ for Model 4A, and $a_{*}=0.97^{+0.02}_{-0.03}$ for Model 4B, within a $1 \sigma$ confidence level. Accordingly, we deduced that the compact star of AT2019 is a close-to-extreme spinning black hole with a dimensionless spin parameter $a_{*}$ $\sim$ $0.96$--$0.97$ and a low inclination angle of the inner disk $i <30 ^{\circ}$. As a caveat, our results are primarily based on \emph{NuSTAR} data alone. Including the lower energy range in addition to the \emph{NuSTAR} data, the column density will be better constrained. However, as discussed in Section \ref{section:4.2}, the column density has a negligible effect on the spin. Furthermore, on the basis of our adopted models, we concluded that AT2019 has a large ionization state of disk $\log \xi\sim$ 3.4--3.6 and a super-solar iron abundance $A_{\mathrm{Fe}} \sim$ 5 $A_{\mathrm{Fe}, \odot}$. In future research, the development of the reflection models will provide additional insight into the properties of black hole systems.

\vspace*{2mm} \Acknowledgements{\bahao We thank the anonymous referees for helpful comments. Ye Feng thank the useful discussions with Prof. Mihoko Yukita on extracting \emph{NuSTAR} spectrum. Ye Feng also thank the useful discussions with Yunfeng Chen. The software used is provided by the High Energy Astrophysics Science Archive Research Centre (HEASARC), which is a service of the Astrophysics Science Division at NASA/GSFC and the High Energy Astrophysics Division of the Smithsonian Astrophysical Observatory. This research has made use of the \emph{NuSTAR} Data Analysis Software (NuSTARDAS) jointly developed by the ASI Space Science Data Center (SSDC, Italy) and the California Institute of Technology (Caltech, USA). Lijun Gou is supported by the National Program on Key Research and Development Project (Grant No. 2016YFA0400804), and by the National Natural Science Foundation of China (Grant No. U1838114), and by the Strategic Priority Research Program of the Chinese Academy of Sciences (Grant No. XDB23040100). }






\end{document}